\documentclass[%
floatfix,
 amsmath,amssymb,
 aps,
 prl,twocolumn
]{revtex4-1}

\usepackage{graphicx}
\usepackage{dcolumn}
\usepackage{bm}

\usepackage{graphicx}
\usepackage{dcolumn}
\usepackage{bm}
\usepackage{float}
\usepackage{physics}
\usepackage{upgreek}
\usepackage{soul}

\begin{document}
\setlength{\abovedisplayskip}{6pt}
\setlength{\belowdisplayskip}{6pt}
\title{Quantum acoustic Fano interference of surface phonons}
\author{J.~M.~Kitzman$^1$}
\email[]{kitzmanj@msu.edu}
\author{J.~R.~Lane$^{1}$}
\author{C.~Undershute$^1$}
\author{N.~R.~Beysengulov$^{1}$}
\author{C.~A.~Mikolas$^{1}$}
\author{K.~W.~Murch$^{2}$}
\author{J.~Pollanen$^{1}$} 
\email[]{pollanen@msu.edu}

\affiliation{$^1$Department of Physics and Astronomy, Michigan State University, East Lansing, MI 48824, USA}
\affiliation{$^2$Department of Physics, Washington University in St. Louis, St. Louis, MO 63130, USA}

\begin{abstract}
    Quantum acoustic systems, which integrate surface or bulk phonons with superconducting qubits, offer a unique opportunity to investigate phononic \emph{interference} and \emph{scattering} processes in the quantum regime. In particular the interaction between a superconducting qubit and a phononic oscillator allows the qubit to sense the oscillator's excitation spectrum and underlying interference effects. Here we present measurements revealing Fano interference of a resonantly trapped piezoelectric surface acoustic wave (SAW) mode with a broad continuum of surface phonons in a system consisting of a SAW resonator coupled to a superconducting qubit. The experiments highlight the existence of additional weakly coupled mechanical modes and their influence on the qubit-phonon interaction and underscore the importance of phononic interference in quantum acoustic architectures that have been proposed for quantum information processing applications.
\end{abstract}

\maketitle
\newpage
Wave interference is a universal phenomenon manifesting in a wide variety of both classical and quantum systems ranging from ocean waves to quantum circuits. The spectral response of these systems encodes the existence of the underlying interference processes, resonant modes, and their losses. A hallmark example is the Fano resonance~\cite{Majorana_1931,PhysRev.124.1866,fanoreview}, which arises from the interference between a resonantly scattered mode and a continuum of background states, and leads to a characteristically asymmetric spectral lineshape. Fano interference has been realized in various fundamentally different quantum systems in which sharp resonant modes interact with continuum excitations, including atomic and molecular systems~\cite{doi:10.1126/science.1234407,PhysRevLett.129.163201}, scattering in optical  experiments~\cite{PhysRevLett.100.043903,Zhao:15}, and transport measurements in quantum dot-based condensed matter systems ~\cite{PhysRevB.70.035319,PhysRevLett.93.106803}. Regardless of the physical platform, this type of interference manifests as a significant change in the spectral response of the system, and is therefore an important process to understand when interpreting spectroscopic or temporal measurements or assessing device performance. Here we demonstrate, for the first time, the manifestation of Fano-interference in a quantum acoustics system. This interference arises from the interaction between resonantly trapped surface acoustic wave (SAW) phonons with a background of continuous phonon modes in an acoustic Fabry-P\'{e}rot resonator that is cooled to near its quantum mechanical ground state. We infer the SAW phonon interference by measuring the absorption spectrum of a superconducting transmon qubit capacitively coupled to the SAW device.
\begin{figure}[b]
    \centering
    \includegraphics[width = 8cm]{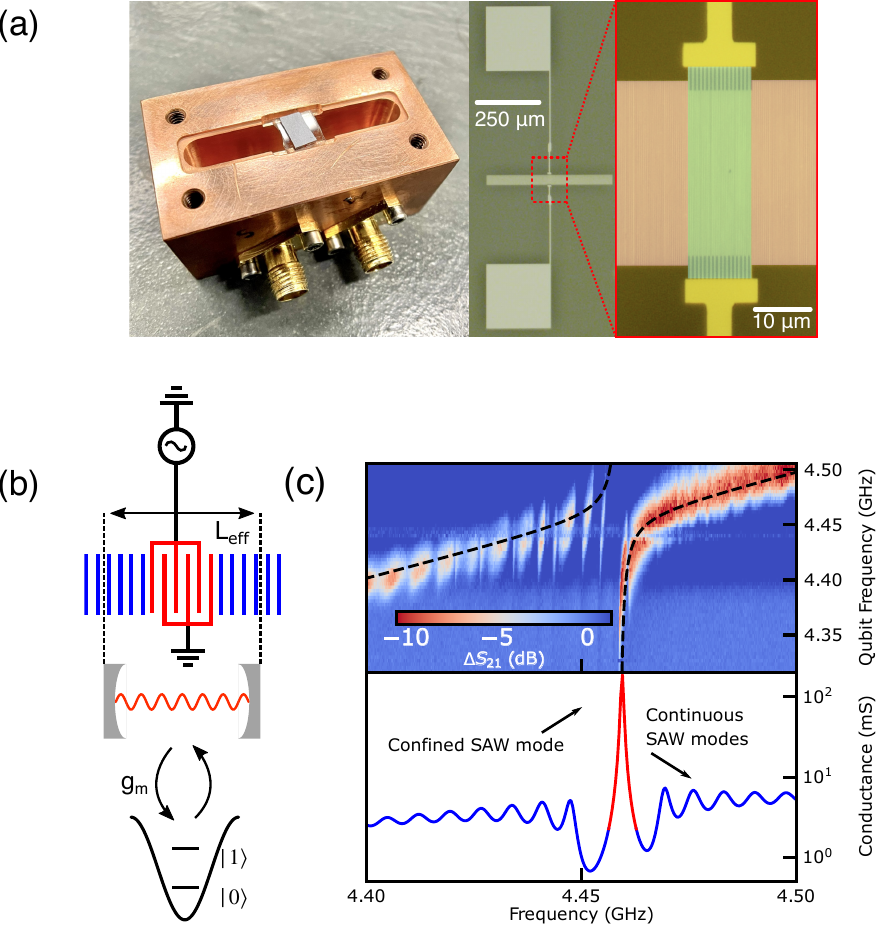}
    \caption{(a) Left: Image of the hybrid system housed in the 3D microwave cavity used for control and readout. Center: Optical micrograph of the SAW resonator. The large antenna mediate the coupling between the qubit and SAW resonator. Right: The acoustic transducer (green) is electrically connected to the antenna pads and phonons are confined within the acoustic cavity by the Bragg mirrors (red). (b) Schematic of the experiment. Phononic Bragg mirrors form a spatially distributed Fabry-P\'{e}rot cavity that allows for a coherent exchange of energy between the qubit and a resonantly confined SAW mode. (c) Top panel: Measurement of two-tone qubit spectroscopy as the qubit is tuned through the main acoustic resonance and nearby SAW continuum. Strong coupling is seen between the confined acoustic mode while weaker coupling is seen between the qubit and the background of acoustic states. Bottom panel: Simulated mode structure for the SAW device, in which the confined acoustic mode is housed within a continuum of SAW states.}
    \label{f1}
\end{figure}

\indent Hybrid quantum systems utilizing the toolkit of circuit quantum electrodynamics (cQED)~\cite{bla04,bla21} are an established and powerful platform for controlling the quantum mechanical properties of microwave photons. These systems have been used to reveal individual photon number states in an microwave field~\cite{gam06,schu07}, to create superpositions of microwave coherent states for generating logical qubits~\cite{vla13,Ofek2016,Gertler2021}, and for creating highly interacting photonic bound states~\cite{Morvan2022}. As a complementary technology, circuit quantum acoustodynamics (cQAD)~\cite{man17}, in which the \textit{photonic} degree of freedom is replaced with a \textit{phononic} one, enabling investigation and control of the quantum properties of vibrational excitations in solid state systems. Experiments using the cQAD framework have been used to investigate the single phonon splitting of the absorption spectrum of a superconducting qubit~\cite{aar19,sle19,vonLupke2022}, the joint entanglement of high-frequency mechanical oscillators~\cite{Wollack2022}, sensing of mechanical dissipation and dephasing~\cite{AYCleland2023}, as well as phononic open quantum systems~\cite{Kitzman2023_be}. While a variety of quantum acoustic systems are available, architectures based on confined or itinerant piezoelectric surface acoustic waves (SAWs) show particular promise for linear mechanical quantum computing~\cite{Qiao2023}, phononic quantum communication~\cite{Bienfait2019,Dumur2021}, and proposed platforms for quantum random access memories~\cite{Hann2019}. In each of these potential quantum information processing applications the presence of phononic interference can lead to a modified interaction between the qubit and mechanical modes of interest, potentially degrading device performance. Our experiments revealing quantum acoustic Fano interference of SAW phonons highlight the importance of understanding the interaction between distinct phonon modes in quantum acoustic systems.

\begin{figure}[b]
    \centering
    \includegraphics[width = 8cm]{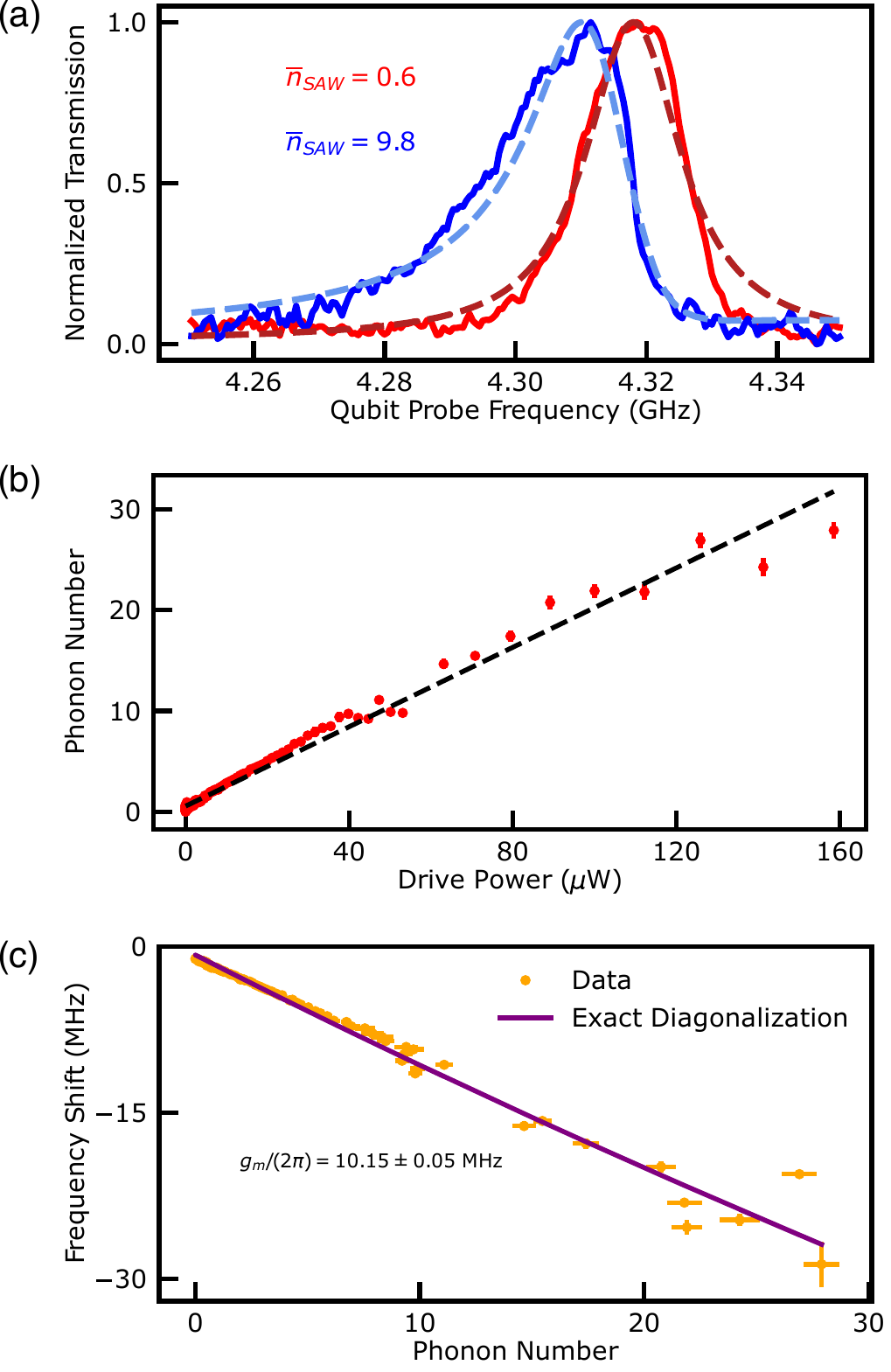}
    \caption{(a) Representative qubit spectra at two different applied powers near resonance with the confined SAW mode. With increasing power the qubit spectrum both shifts in frequency and inherits a spectral lineshape expected from the statistics of the acoustic coherent state. (b) Measurement of mean phonon number as a function of drive power. The extracted phonon number follows the expected linear trend (black). (c) Qubit frequency shift versus extracted mean phonon number. Exact diagonalization considers the qubit as a 5 level atom, which agrees with the experiment.}
    \label{f2}
\end{figure}
\indent Figure~\ref{f1}a depicts a schematic of our experiment, which consists of a flux-tunable superconducting transmon qubit that is capacitively coupled with a SAW device hosting a resonant mode at $\omega_m/(2\pi) = 4.4588$~GHz. The spectral response of the SAW device was precisely designed using the coupling-of-modes theory to define the electro-mechanical scattering properties of the device~\cite{mor05book,lane21thesis}, as shown in Fig.~\ref{f1}b. The effective electrical conductance of the SAW device was designed such that the resonantly confined acoustic mode is in close spectral proximity to a continuum of acoustic states (see Fig.~\ref{f1}b and Appendix A). The qubit and SAW device are fabricated on separate substrates, with the SAW device on \textit{YZ}-cut LiNbO$_3$ and the qubit on high-resistivity silicon with maximum Josephson tunneling energy $E_{J,\textrm{max}}/h = 19.7$~GHz and capacitive charging energy $E_C/h = 328$~MHz. Both devices are galvanically connected to large antenna pads having an area $250~\mu \textrm{m} \times 250~\mu \textrm{m}$ that form a pair of parallel plate capacitors between the two devices when they are assembled in a flip-chip configuration~\cite{device}. The capacitive coupling in this hybrid system ensures that strain in the piezoelectric SAW substrate induces a voltage across the qubit antennae, allowing for the exchange of energy between the qubit and SAW phonons (see Fig.~\ref{f1}a). The composite qubit-SAW system is housed in a 3D electromagnetic cavity with fundamental frequency $\omega_c/(2\pi) = 4.788$~GHz, which is used for qubit control and readout as well as for applying independent excitation tones to populate the SAW device, and the interaction strength between the qubit and microwave cavity is measured to be $g/(2\pi) = 75$~MHz. A superconducting coil wound around the cavity provides magnetic flux tunability of the resonant frequency of the qubit. The qubit absorption spectrum is measured via two-tone spectroscopy, using the dispersive interaction between the qubit and cavity to determine the qubit state. As the resonant frequency of the qubit is tuned through the confined acoustic mode, we observe an avoided crossing of magnitude $g_m/(2\pi) = 9.76\pm0.60$~MHz (see Fig.~\ref{f1}b). Additional interactions between the qubit and phonon modes are also observed. In particular these features correspond to interactions between the qubit and phonons that are not strongly confined to the SAW resonator and therefore couple much more weakly to the qubit, manifesting as a series of dark states in the qubit spectra.

To utilize the qubit to measure surface phonon excitations, we first calibrate the qubit response to SAW excitations in the acoustic dispersive limit using the primary confined SAW resonance. In this regime the qubit frequency and spectral shape depend in an established and systematic fashion on the excitation number of SAW bosons. This ac Stark shift has been observed in superconducting qubit systems coupled to both microwave resonators in the cQED framework~\cite{schu05,schu07} as well as mechanical oscillators of multiple cQAD architectures~\cite{man17,sle19,aar19,vonLupke2022}.  In this dispersive limit, the detuning between the qubit frequency ($\omega_q$) and the resonant SAW mode ($\omega_m$) is large compared to $g_m$ ($g_m \ll |\Delta|$, $\Delta = \omega_q - \omega_m$). In particular we tune the qubit frequency such that $\Delta/(2\pi) = -138.6$~MHz and $E_J/h = 8.5$~GHz. In this regime, we can approximate the Hamiltonian describing the hybrid system as~($\hbar = 1$)~\cite{bla04}
\begin{equation}\label{eq1}
    \hat{H} \simeq \omega_{m} \left(\hat{a}^{\dagger} \hat{a} + 1/2 \right) + \frac{1}{2 }\left( \omega_q + 
   2\chi_m \hat{a}^{\dagger}\hat{a} + \frac{g_m^2}{\Delta}\right) \hat{\sigma}_z.
\end{equation}
\noindent
In Eq.~(\ref{eq1}) the SAW degrees of freedom are described by bosonic operators $\hat{a}$ and $\hat{a}^{\dagger}$, and the qubit is described by the spin 1/2 operator $\hat{\sigma}_z$. By considering the transmon as a multilevel artificial atom, the total frequency shift of the qubit frequency per piezophonon, $2\chi_m$ must take into account multiple partial dispersive shifts and is therefore given by~\cite{koc07}
\begin{equation}
    2\chi_m = -\frac{2g_m^2}{\Delta}\frac{\alpha}{\Delta - \alpha},
\end{equation}
where $\alpha/h = 328$~MHz is the anharmonicity of the transmon.
Based on the experimental parameters of our system the qubit frequency shift per phonon at this detuning is $2\chi_m/2\pi = -0.97$~MHz. 

By applying a tone that is resonant with the confined acoustic mode we generate a coherent SAW state and subsequent measurement of the resulting qubit spectra allows us to determine the mean SAW resonator occupancy number. We fit the qubit spectra (see Fig.~\ref{f2}a) to a model~\cite{gam06} consisting of a two level system coupled to a harmonic oscillator coherent state and extract the average phonon occupation number $\overline{n}$. The asymmetry of the qubit spectra at large excitation numbers (see Fig.~\ref{f2}a) arises from the statistics of the distribution of phonon numbers in the resulting SAW coherent state. As shown in Fig.~\ref{f2}b we find a linear relationship between the drive power populating the SAW resonator and the extracted mean phonon number. We note that the power reported in our measurements is that applied at room temperature, which is further attentuated by 60~dB in the cryostat before entering the microwave cavity. Exact diagonalization of a multi-level Jaynes-Cummings Hamiltonian describing this coupled system allows us to extract the expected qubit frequency shift as a function of phonon number in the dispersive limit~\cite{PhysRevLett.120.227701} and we find excellent agreement between this prediction and the measured phonon number as shown in Fig.~\ref{f2}c. In this analysis the coupling strength $g_{m}$ is a fit parameter to the data and we extract $g_m/(2\pi) = 10.15 \pm 0.05$~MHz, which is in reasonable agreement with the measured value determined from fitting the avoided crossing between the qubit and SAW modes.
\begin{figure}[b]
    \centering
    \includegraphics[width = 8cm]{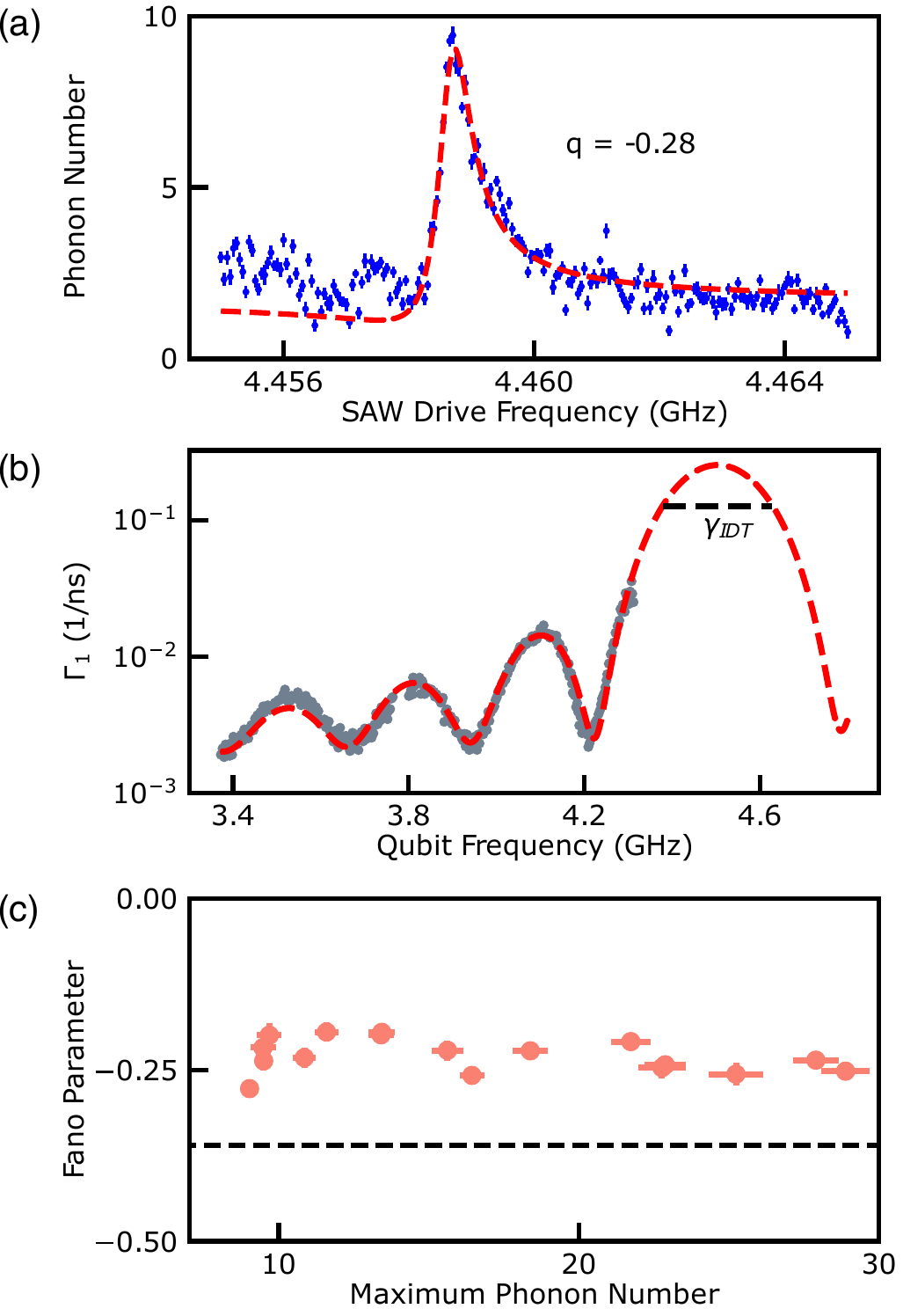}
    \caption{ (a) Mean phonon number as a function of SAW drive frequency at a fixed power of $\textrm{P}_{\textrm{drive}} = 25~\mu \textrm{W}$. The SAW resonance is asymmetric in frequency and is well-described by a Fano absorption function. The asymmetry arises from phonons in the confined acoustic mode interacting with phonons in the SAW continuum. (b) Measurement of the qubit decay rate over a broad range of frequencies far from the confined acoustic resonance. The fit (red) is to the analytical expression of the SAW-induced loss due to the central SAW transducer. The decay rate of the transducer is $\gamma_{\textrm{IDT}}/(2\pi) = 249.7$~MHz. (c) Fano asymmetry parameter $q$ as a function of drive power. The horizontal error bars correspond to the uncertainty in the fit parameter ${\overline{n}_{\textrm{max}}}$ in Eq.~(\ref{fano}). The black dashed line indicates $q$ based on the analysis of two coupled harmonic oscillators as described in the text.} 
    \label{f3}
\end{figure}
Having calibrated the frequency response of the qubit to phonons in the confined SAW resonance, we extend these Stark shift measurements to probe the acoustic environment outside of the stop-band of the SAW mirrors. We measure the acoustic Stark shift $\delta \omega_q$ as a function of drive frequency $\omega_d$, and extract the mean phonon occupation $\overline{n} = \delta \omega_q/2\chi_m$ over a range of drive frequencies. As shown in Fig.~\ref{f3}a, the measured phonon number has a maximum at the resonant frequency of the confined acoustic mode $\omega_m/(2\pi) = 4.4588$~GHz, and is strikingly asymmetric about this peak. The acoustic excitation spectrum hosts a rich structure near this confined resonance, making it possible for phonons across a range of frequencies to interfere with each other. Because the reflectivity of the mirrors that define the SAW cavity is relatively low ($\sim 0.5\%$ per mirror structure), the reflection process for surface phonons is distributed over the length of the cavity. This creates a situation in which phononic excitations that reflect at different spatial positions within the cavity interfere with each other either constructively or destructively depending upon their relative wavevectors and propagation paths. These interference processes produce a Fano lineshape, as phonons in the confined acoustic mode are subject to differing interference with continuum surface phonons existing outside of the mirror stop-band. We model the resulting frequency-dependent phonon number as a Fano resonance~\cite{PhysRev.124.1866,fanoreview} with an absorption spectra, $\overline{n}(\omega)$:
\begin{equation}\label{fano}
    \hspace{-0.5mm}\overline{n}(\omega) = {\overline{n}_{\textrm{max}}} \left(1+q^2-\frac{(q\Gamma/2+\omega-\omega_m)^2}{(\Gamma/2)^2+(\omega-\omega_m)^2}\right)+\overline{n}_{\textrm{off}}.
\end{equation}
\noindent Equation~(\ref{fano}) depends on the linewidth $\Gamma$, of the resonantly confined mode, the average background population of the continuum phonon modes $\overline{n}_{\textrm{off}}$, as well as the Fano parameter $q$, which describes the level of interference between the confined and lossy surface phonons. The fit parameter ${\overline{n}_{\textrm{max}}}$ quantifies the maximum phonon number for a given measurement, which is set by the drive power. The first two terms in Eq.~(\ref{fano}) do not depend on frequency and impose the physical constraint that the minimum SAW phonon number is non-negative. We note that the limit $q~\xrightarrow{}~0$ corresponds to the absence of phonon interference, and in this limit a Lorentzian response is recovered.

To determine the characteristic level of phonon interference in our device, we use the qubit to measure the phonon occupation near the confined acoustic mode as a function of the power used to populate the SAW device with phonons. By fitting each resulting phonon spectra using Eq.~(\ref{fano}) we determine the Fano parameter as a function of the maximum mean phonon number. These results provide a relative measure of the acoustic interference in the device and are displayed in Fig.~\ref{f3}c. In particular, we find that the phonon interactions are well-described by Eq.~(\ref{fano}), with a negative Fano parameter, down to the lowest phonon levels we are able to measure. Additionally, we find that the phonon interference is roughly constant with $q \simeq -0.25$.

To understand the interference we observe in our measurements, we compare the SAW phononic system to a minimal classical model of Fano interference arising in two coupled oscillators in which one oscillator has a significantly larger loss rate than the other~\cite{Iizawa_2021}. In this model, the lossy oscillator approximates a continuum over the frequency scale of the confined mode (see Appendix A) and the interaction between the two oscillators leads to Fano interference. The loss rate of the confined acoustic mode, which corresponds to the low loss oscillator, is extracted from the linewidth of the fit to the resonantly trapped SAW mode in Fig.~\ref{f3}a. This linewidth indicates that the loss of the confined mode is $\gamma_{\textrm{SAW}}/(2\pi) = 630$~kHz, corresponding to a SAW quality factor $Q_{\textrm{SAW}} \simeq 7000$. To estimate the loss rate of the significantly broader interdigitated transducer (IDT) response, which acts as an effective continuum over the scale of $\gamma_{\textrm{SAW}}$, we measure the qubit decay rate, $\Gamma_1 = 1/T_1$, as a function of qubit frequency far from the confined acoustic resonance. In this regime, the qubit loss is proportional to the electrical conductance~\cite{Houck2008,Kitzman2023_be} of the SAW device, which is well described by only the IDT response far from the confined mode, where the mirror reflectivity is much less than one. In particular we fit $\Gamma_1$ to a phenomenological form taking into account loss resulting in the transduction of qubit excitations into phonons that exit the SAW mirrors,
\begin{equation}\label{losseq}
    \Gamma_1\left(\omega_q\right) = \frac{\omega_q}{Q_i} + \Gamma_0~\textrm{sinc}^2\left(\pi N_p \frac{\omega_q-\omega_{IDT}}{\omega_{IDT}} \right),
\end{equation}
\begin{figure}[t]
    \centering
    \includegraphics[width = 8cm]{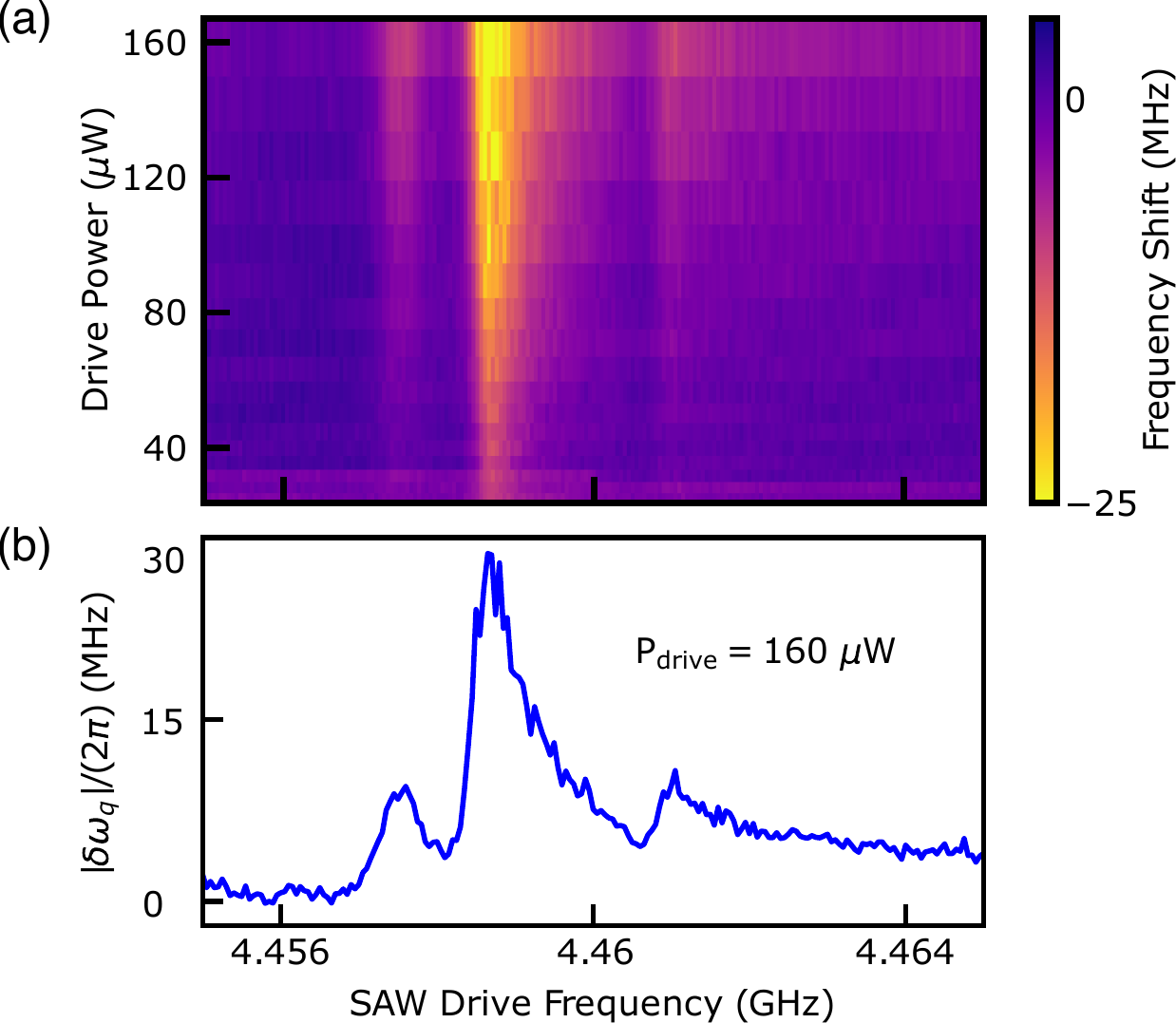}
    \caption{(a) Spectra of the SAW resonator inferred from the phonon-induced qubit Stark shift as a function of power. As the drive power is increased, two additional spectral features surrounding the strongly confined mode are visible, which we attribute to SAW modes that are weakly confined by the Bragg mirrors. (b) Horizontal linecut of panel (a) taken at a drive power of $P_{\textrm{drive}} = 160~\mu$W.}
    \label{f4}
\end{figure}
 where $Q_i = 1.05\times10^4$ is the qubit internal quality factor~\cite{quality_factor}, $\Gamma_0 = 0.252~$ns$^{-1}$ is the maximum conversion rate of the qubit excitation into SAW phonons, $N_p = 16$ is the number of finger pairs in the IDT structure of the SAW resonator, and $\omega_{\textrm{IDT}}/(2\pi) = 4.504$~GHz is the central transducer frequency, which is within 1\% of the value predicted from the device fabrication parameters. By fitting the qubit loss to the response given by Eq.~(\ref{losseq}) (see Fig.~\ref{f3}b), we are able to extract the overall loss associated with the central transducer, which we approximate as a Lorentzian with a width of $\gamma_{\textrm{IDT}}/(2\pi) = 249.7$~MHz. In the limit where the response of the transducers is broad in frequency compared to that of the confined mode, as is the case in our experiments ($\gamma_{\textrm{SAW}}/\gamma_{\textrm{IDT}} \approx 0.3\%$), the Fano parameter can be calculated as~\cite{Iizawa_2021},
\begin{equation}\label{hoq}
    q = \frac{1}{\gamma_{\textrm{IDT}}~\omega_{\textrm{SAW}}}\left(\omega_{\textrm{SAW}}^2 - \omega_{\textrm{IDT}}^2\right).
\end{equation}
Based on the experimental parameters this model predicts $q = -0.36$, which is indicated by the black dashed line in Fig.~\ref{f3}b. The systematic deviation between the experimentally determined and predicted values of $q$ could arise from the presence of additional mechanical modes near the confined SAW resonance. In fact, in our higher power measurements shown in Fig.~\ref{f4}(a,b) we observe evidence for the existence of such modes on either side of the strongly confined SAW resonance. These modes also exhibit asymmetry, indicating that they are also interfering with the continuous phonon background. Since the calculated free spectral range of the SAW resonator is comparable to the width of the mirror stop-band, it is likely that these additional modes correspond to surface acoustic waves weakly confined within the SAW cavity (see Appendix A).

In conclusion, we have demonstrated the Fano interference of surface acoustic wave phonons hybrid quantum acoustic device containing a resonantly confined SAW mode embedded in a continuum of surface phonons. This phononic interference is inferred from qubit-assisted spectroscopy of the SAW device and we find that it persists down to extremely low excitation number. The experimental results are in excellent agreement with the functional form of a Fano resonance and highlight the importance of phononic interference in quantum acoustic devices proposed for applications in quantum information processing.

We thank M.I.~Dykman, V.~Zelevinsky, A.~Schleusner, D.~Kowsari, P.M.~Harrington, and P.K.~Rath for valuable discussions. We also thank R.~Loloee and B.~Bi for technical assistance and use of the W.~M.~Keck Microfabrication Facility at MSU. The Michigan State portion of this work was supported by the National Science Foundation (NSF) via grant number ECCS-2142846 (CAREER) and the Cowen Family Endowment. The Washington University portion of this work was also supported by the NSF via grant number PHY-1752844 (CAREER).

\section{Appendix A: SAW Resonator Design}
\begin{figure}[h]
    \centering
    \includegraphics[width = 8cm]{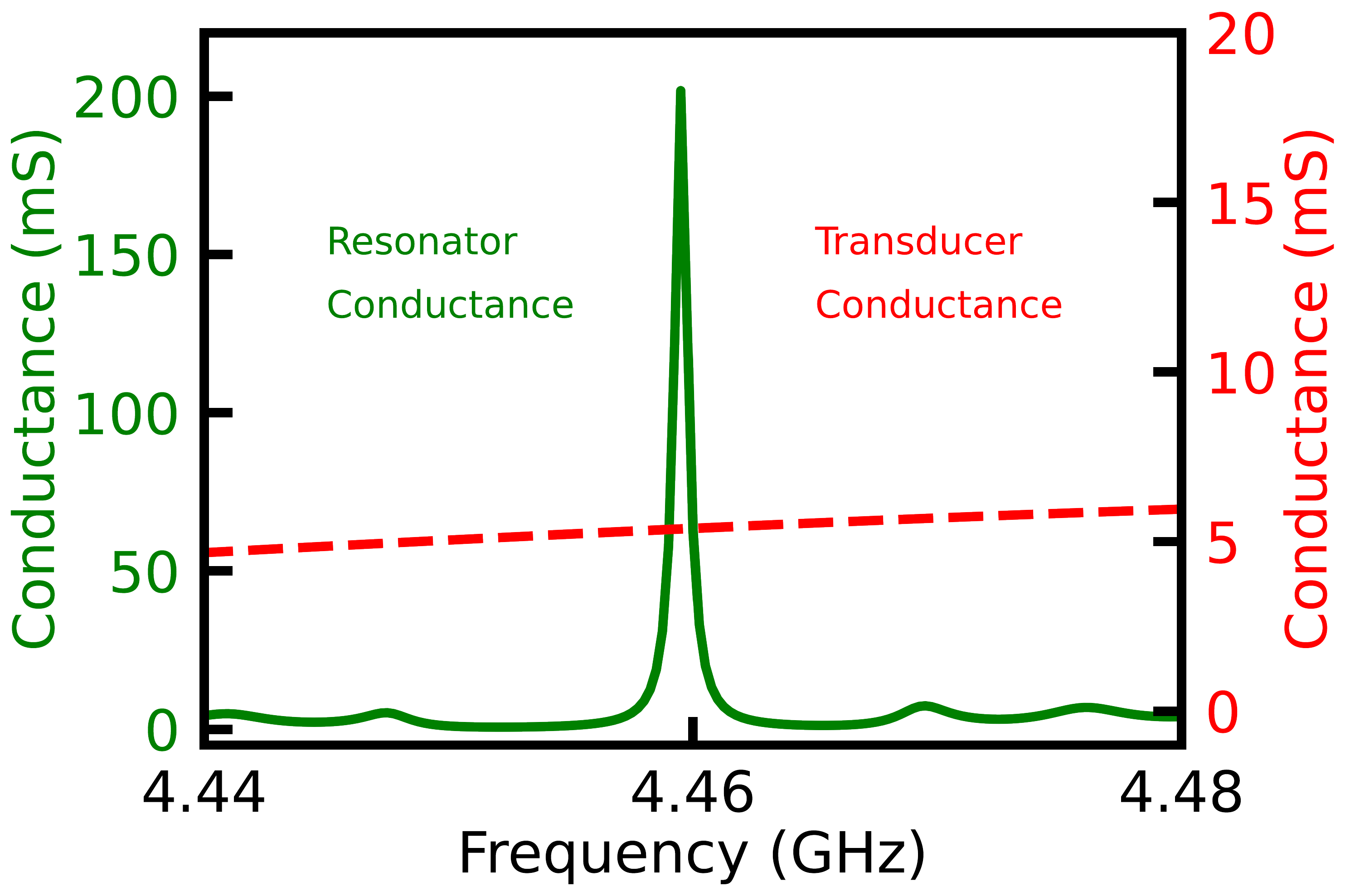}
    \caption{Solid green curve: Simulated conductance of the phononic resonator consisting of a central acoustic transducer enclosed in Bragg mirrors. Dashed red curve: simulated conductance of the acoustic transducer alone. The linewidth of the central resonance is much smaller than the width of the transducer response, which can be approximated as a continuous background.}
    \label{f5}
\end{figure}

The spectral response of the SAW resonator was calculated using coupling-of-modes (COM)~\cite{lane21thesis}. A complete list of the SAW device parameters is presented in Table~1. Since the reflectivity of the mirrors is relatively low per mirror grating, on average acoustic waves penetrate a distance $L_P \simeq \lambda_{\textrm{mirror}}/2|r_m|$ into the mirrors before being reflected. Based on the parameters in Table 1, we calculate $L_P =  81.6~\mu m$ and the total effective length of the phonon cavity is $L_{\textrm{eff}} = L_{\textrm{IDT}} + 2L_P = 175.2~\mu m$. The free spectral range of the SAW resonator can then be estimated as $\Delta f_{\textrm{FSR}} = v_{\textrm{s}}/2L_{\textrm{eff}} = 10.4~\textrm{MHz}$. The width of the mirror stop band is given by $\Delta f_{\textrm{mirror}} = \frac{2|r_m|f}{\pi} = 14.2~\textrm{MHz}$~\cite{mor05book,lane21thesis}. Since the width of the mirror stop band is comparable to the mode spacing, we expect the resonator to host a single strongly confined acoustic mode as well as other more weakly reflected SAW modes as seen in Fig.~\ref{f4} and \ref{f5}.
\begin{table}[t]
 \centering
 \begin{tabular}{||c c c||} 
 \hline
 Parameter & Physical quantity & Value\\
 \hline\hline
     $\lambda_{\textrm{IDT}}$ & Transducer periodicity & $800~$nm\\ 
     \hline
     $\lambda_{\textrm{mirror}}$ & Mirror periodicity & $816~$nm\\
     \hline
     $N_P$ & Number of finger pairs&16\\
     \hline
     W & Finger pair overlap & 35$~\mu$m\\
     \hline
     $L_{\textrm{mirror}}$ & Bragg mirror length & 240.72$~\mu$m\\
     \hline
     $L_{\textrm{IDT}}$ & Transducer length & 12$~\mu$m\\
     \hline
     $v_{\textrm{s}}$ & Speed of sound & $3638~$m/s\\
     \hline
     $\eta$ & SAW propagation loss & 500 Np/m\\
     \hline
     $r_i$ & Transducer reflectivity & -0.005i\\
     \hline
     $r_m$ & Mirror reflectivity & -0.005i\\
\hline
\end{tabular}
\caption{Summary of SAW device parameters.}
\label{table}
\end{table}
In Fig.~\ref{f5} we plot the simulated conductance of the composite resonator along with the conductance of the IDT structure. On the frequency scale of the confined mode ($\sim 1$~MHz) the IDT response is approximately constant creating a continuum of background SAW phonons.
%


\begin{thebibliography}{37}
\bibitem [{qua()}]{quality_factor}%
  \BibitemOpen
  \href@noop {} {}\bibinfo {note} {The internal qubit quality factor is defined
  as $Q_i \equiv 2\pi \times f_q \times T_{1i}$, where $f_q$ is the qubit
  frequency and $T_{1i}$ is the intrinsic lifetime of the qubit. We note that
  this differs from the definition used in Ref.~\cite{Kitzman2023_be}, in which
  $Q_i \equiv f_q \times T_{1i}$}\BibitemShut {NoStop}%
\end{thebibliography}
\end{document}